\documentclass[aps,prb,preprint,amsmath,amssymb]{revtex4-1}
\usepackage{graphicx}
\usepackage{dcolumn}
\usepackage{hyperref}
\begin{document}
\title{Negative longitudinal magnetoresistance \\as a sign of a possible chiral magnetic anomaly \\in the half-Heusler antiferromagnet DyPdBi}
\author{Orest Pavlosiuk} 		
\author{Dariusz Kaczorowski}	
\author{Piotr Wi{\'s}niewski}    
\email[Corresponding author: ] {p.wisniewski@intibs.pl}
\affiliation{Institute of Low Temperature and Structure Research,
Polish Academy of Sciences, P.\,Nr 1410, 50-950 Wroc{\l}aw, Poland}
\date{\today}
\begin{abstract}
Magnetotransport investigation of a half-Heusler antiferromagnet DyPdBi revealed  hallmark features of Weyl semimetal: huge negative longitudinal magnetoresistance and planar Hall effect. Both effects have recently been linked to chiral magnetic anomaly - axial charge pumping between Weyl nodes.  
Magnetoresistance (MR) of single crystals of DyPdBi is very pronounced. In magnetic field longitudinal to electrical current direction it reaches -80\% and its relative difference with respect to that measured in transverse field (expressed as anisotropic magnetoresistance) is extremely strong: -60\% at 10\,K and 14\,T.
The planar Hall effect in DyPdBi depends on temperature and magnetic field in non-monotonous way, which has not been previously reported.
We compare magnetoresistance measured with voltage contacts on mid-line of the sample with that measured with contacts on its edge, and show that the role of current-jetting, an extrinsic source of anisotropic negative magnetoresistance, is marginal.
We discuss that nature of the compound and sample quality exclude intrinsic sources of negative and anisotropic magnetoresistance other than weak localization and the chiral magnetic anomaly.
\end{abstract}
\maketitle 
\section{Introduction}
Topological semimetals have exotic quantum mechanical phenomena at easily accessible temperatures and magnetic fields, which makes them good candidates for novel technological applications (quantum computing, spintronics, etc.) \cite{Hasan2017,Bernevig2018,Armitage2018}.  
Weyl semimetals (WSMs) belonging to this group are distinguished by the presence of Weyl nodes, contact points between bands, always occurring in pairs with opposite chirality and behaving like magnetic monopoles \cite{Nielsen1983,Wan2011}. Low-energy excitations in WSM are massless Weyl fermions, surface states forming Fermi-arc in momentum space. 

Unusual electronic structure of WSMs gives rise to peculiar phenomenon, the chiral magnetic anomaly (CMA). CMA is the charge pumping between lowest Landau levels of two Weyl nodes of opposite chirality, which arises when strong magnetic field, ${\bf B}$, is applied along electric current in a sample, and is manifested by negative longitudinal magnetoresistance (nLMR) \cite{Nielsen1983,Son2013}. 

Further consequence of CMA is the anisotropic magnetoresistance (AMR), i.e. angular dependence of diagonal term of resistivity tensor, $\rho_{xx}\propto V_x/j_x$, when field direction is rotated in $x\!-\!y$ plane, and $V_x$ is voltage drop measured along the current, $j_x$, flowing in direction $x$. Simultaneously, the off-diagonal component of that tensor, $\rho_{yx}\propto V_y/j_x$, is also field-angle dependent, but in a different manner. This is traditionally called planar Hall effect (PHE), because the voltage drop, $V_y$, is here measured transverse to the current. In contrast to conventional Hall effect, PHE is even in magnetic field. 

Both AMR and PHE are due to enhancement of conductance by  component of magnetic field parallel to the current. Therefore, nLMR, AMR and PHE have been hailed as fingerprints of Weyl semimetals \cite{Burkov2017,Nandy2017}.
Accordingly, these phenomena have been observed in topologically nontrivial materials, such as thin films of topological insulator 
Bi$_{2-x}$Sb$_x$Te$_3$ (PHE\cite{Taskin2017c}), 
or topological semimetals:  
Na$_3$Bi (nLMR\cite{Xiong2015a,Liang2018b}), 
ZrTe$_5$ (nLMR \cite{Li2016})
Cd$_3$As$_2$ (PHE \cite{Li2018c}),
TaAs (nLMR \cite{Huang2015,Zhang2016a}), 
WTe$_{1.98}$ (nLMR \cite{Lv2017}), 
YbPtBi (nLMR and PHE \cite{Guo2018a}),
and GdPtBi (nLMR \cite{Hirschberger2016a,Shekhar2016a,Liang2018b} and PHE \cite{Kumar2018a, Liang2018b}). 

However, AMR and PHE, have also been observed in materials with ferromagnetically ordered moments. There AMR occurs due to anisotropic spin-orbit coupling strongly modifying spin-flip inter-band scattering rates of electrons \cite{Coey2009} or magnetostriction-induced modification of Fermi surface \cite{Wisniewski2007}. Also nonmagnetic materials such as monopnictides of Y, La and Lu have recently been shown to exhibit very strong AMR \cite{Pavlosiuk2016d, Pavlosiuk2017a}. So AMR is not a proof but only an indication of possibility of CMA. 

Several mechanisms, such as de Gennes-Friedel spin-disorder scattering and weak localization, can produce negative contributions to magnetoresistance \cite{DeGennes1958,Lu2017b}. However, apart from  CMA only weak localization and spurious current-jetting effect generates negative MR dependent on the direction of applied field \cite{Liang2018b}. 
Therefore, if the current-jetting is eliminated, the appearance of nLMR can indicate the presence of CMA and Weyl fermions in studied material. 

Several half-Heusler (HH) bismuthides bearing rare-earth elements were theoretically predicted to exhibit topologically nontrivial character \cite{Chadov2010a}, which may coexist at low temperatures with superconductivity and/or antiferromagnetic ordering. 
Examples of experimentally characterized superconducting HH putative topological insulators or  semimetals include nonmagnetic: YPtBi, LuPtBi, and LuPdBi \cite{Butch2011a,Tafti2013,Xu2014,Pavlosiuk2015, Pavlosiuk2016b,Liu2016a} as well as antiferromagnetic $R$PdBi ($R=$ Sm, Dy, Ho, Er, Tm)\cite{Pan2013a,Nikitin2015a,Nakajima2015d, Pavlosiuk2016a} phases. 
Recently, GdPtBi has been described as the first WSM with Weyl nodes induced by magnetic field \cite{Hirschberger2016a}, manifesting strong CMA \cite{Hirschberger2016a,Shekhar2016a,Liang2018b} and huge PHE \cite{Kumar2018a}. 

DyPdBi has the same crystal and antiferromagnetic structure as GdPtBi.
Measurements on its polycrystalline samples have shown N{\'e}el temperature, $T_{\rm N}$, of 3.5\,K\ \cite{Gofryk2005}, those performed on single crystals revealed $T_{\rm N}$ = 2.7\,K  \cite{Nakajima2015d} or $T_{\rm N}$ = 3.7\,K \cite{Pavlosiuk2016c}. The magnetic structure of DyPdBi determined by neutron diffraction is characterized by the propagation vector ($^1\!/_2, ^1\!/_2, ^1\!/_2$) \cite{Nakajima2015d,Pavlosiuk2018}, same as for GdPtBi  \cite{Muller2014a}. DyPdBi becomes superconducting below 0.9\,K \cite{Nakajima2015d}.

We carried out comprehensive measurements of electrical transport of single-crystalline DyPdBi with emphasis on properties potentially revealing CMA. 
Obtained nLMR, AMR and PHE data hint at possible topologically nontrivial character and allow to presume that DyPdBi is a Weyl semimetal.

\section{Experimental details}
Single crystals of DyPdBi were grown from Bi flux and characterized as reported before  \cite{Pavlosiuk2016c}. 
Cuboid samples suitable for electrical transport measurements were cut along $\langle 100 \rangle$ crystallographic directions from single crystals oriented using backscattering Laue method (see Supplemental Material \cite{SupMat} for a typical Laue diffraction pattern). Electrical contacts were made with $50\,\mu\!$m silver wire attached to the samples with silver paint. 

Electrical transport measurements were performed using conventional four-probe method in temperature range 10--300\,K and in magnetic fields up to 14\,T employing a Quantum Design PPMS. The experimental data were either symmetrized or antisymmetrized in order to dispose of contact misalignment effect.
\section{Results}
\subsection{Longitudinal and transverse magnetoresistance}
\begin{figure}[h]
	\includegraphics[width=0.49\textwidth]{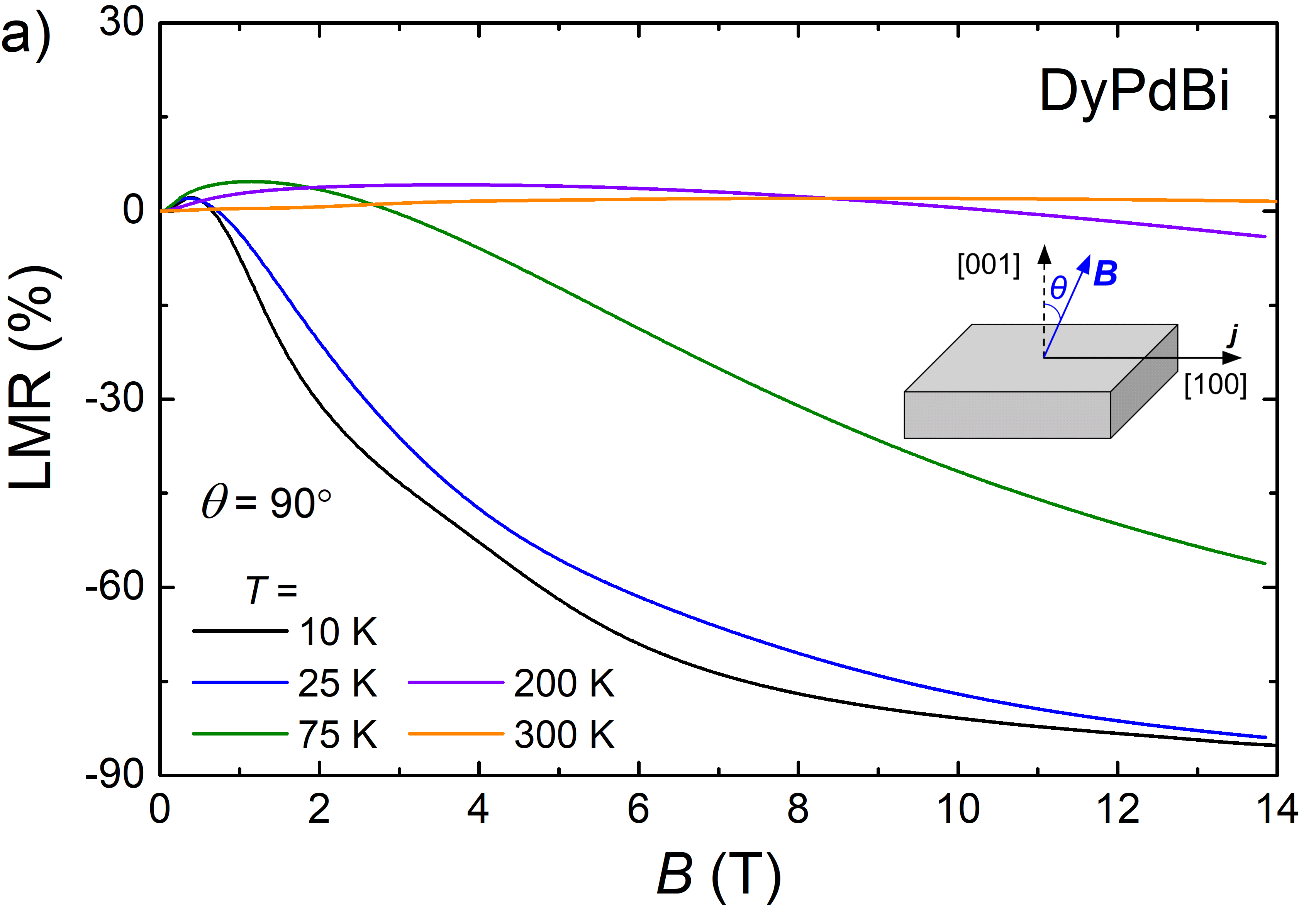}
	\includegraphics[width=0.49\textwidth]{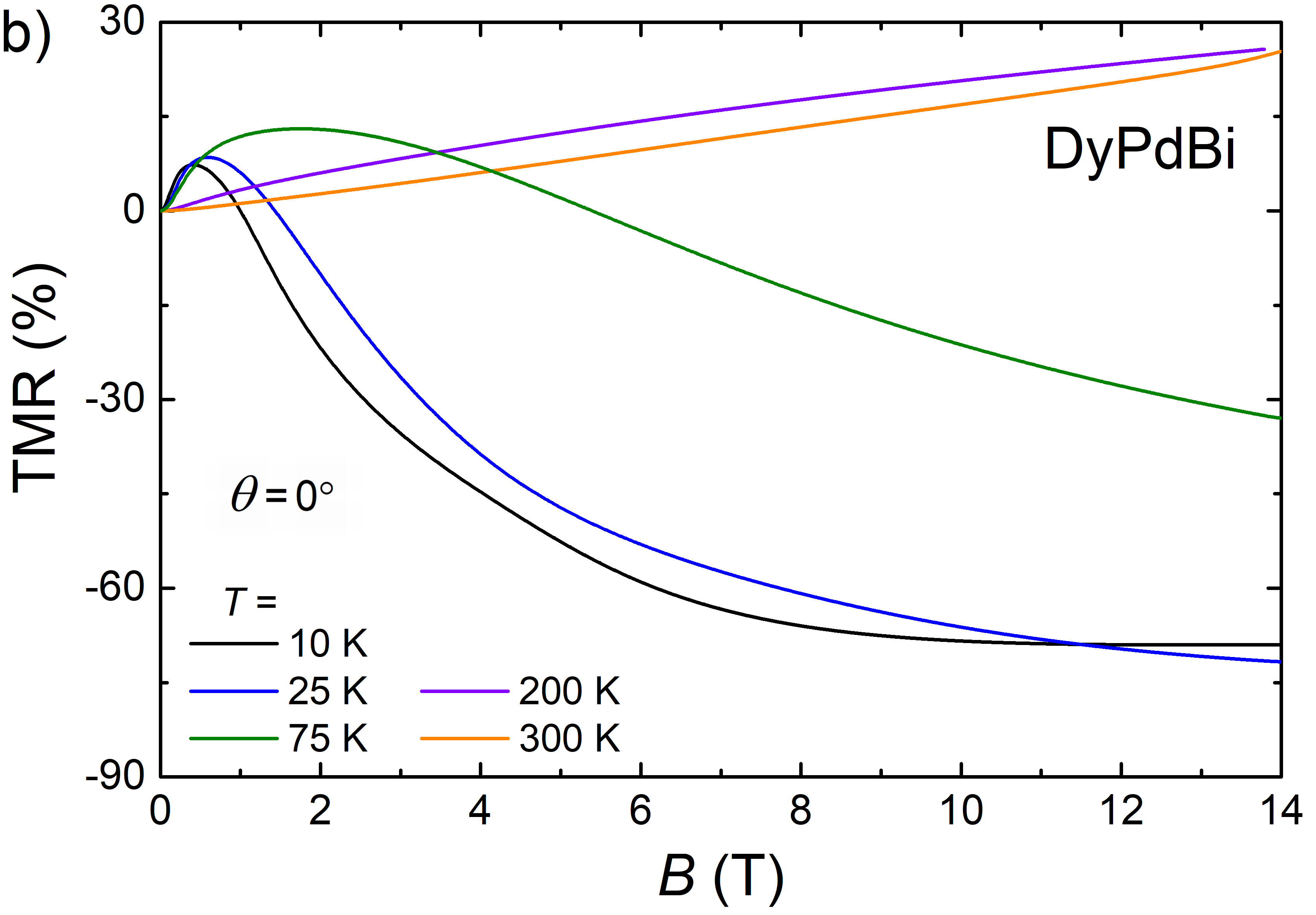}
	\caption{Magnetoresistance of single-crystalline DyPdBi measured as a function of magnetic field strength in temperature range from 10\,K to 300\,K in (a) longitudinal and (b) transverse configuration of electrical current and magnetic field. Inset to panel (a) shows schematically the configuration of measurements.}
	\label{rho}
\end{figure}

Magnetoresistance, MR\,$\equiv\!\rho_{xx}(B)/\rho_{xx}(B\!=\!0)\!-1$, of DyPdBi measured at several temperatures from the range 10--300\,K in longitudinal (LMR; ${\bf B}\parallel{\bf j}\parallel a$-axis) and transverse (TMR; ${\bf B}\perp{\bf j} \parallel a$-axis) configuration, is presented in Figs.~\ref{rho}(a) and \ref{rho}(b), respectively. 

Qualitatively, behavior of LMR and TMR is very similar. 
In weak fields, they rapidly grow with increasing $B$. Similar behavior has been observed in several HH phases and ascribed to weak antilocalization effect \cite{Pavlosiuk2015, Pavlosiuk2016a,Pavlosiuk2016b, Hou2015a,Hou2015b,Bhardwaj2018}. At certain values of $B$, both LMR and TMR attain maximum, then monotonically decrease and become negative.
These field variations become less pronounced with increasing temperature and above certain temperature both LMR and TMR are positive in the entire magnetic field range covered.

Negative LMR in DyPdBi is observed at temperatures up to $T=200$\,K, and in this regard is very similar to that of GdPtBi \cite{Hirschberger2016a}. At $T=10$\,K and in $B=14$\,T it reaches -85\%. This is very close to -83\% recorded at 6\,K for GdPtBi in Ref.\cite{Hirschberger2016a}.
It is worth noting  that absolute limit of negative value of MR is -100\% (such value would mean that magnetic field suppresses the resistivity completely, like in quantum Hall systems), so nLMR found for DyPdBi is really remarkable. 

TMR also has outstanding negative values ($\approx$\,-70\% in $B=$\,14\,T, at $T=$\,10 and 25\,K) in DyPdBi, in clear contrast to the GdPtBi, where TMR at the same fields and temperatures was positive ($\approx$\,200\%) \cite{Hirschberger2016a,Shekhar2016a}. 

Nevertheless, negative LMR is significantly stronger than negative TMR measured at the same temperatures, as discussed below.
\begin{figure*}[h]
	\includegraphics[width=0.49\textwidth]{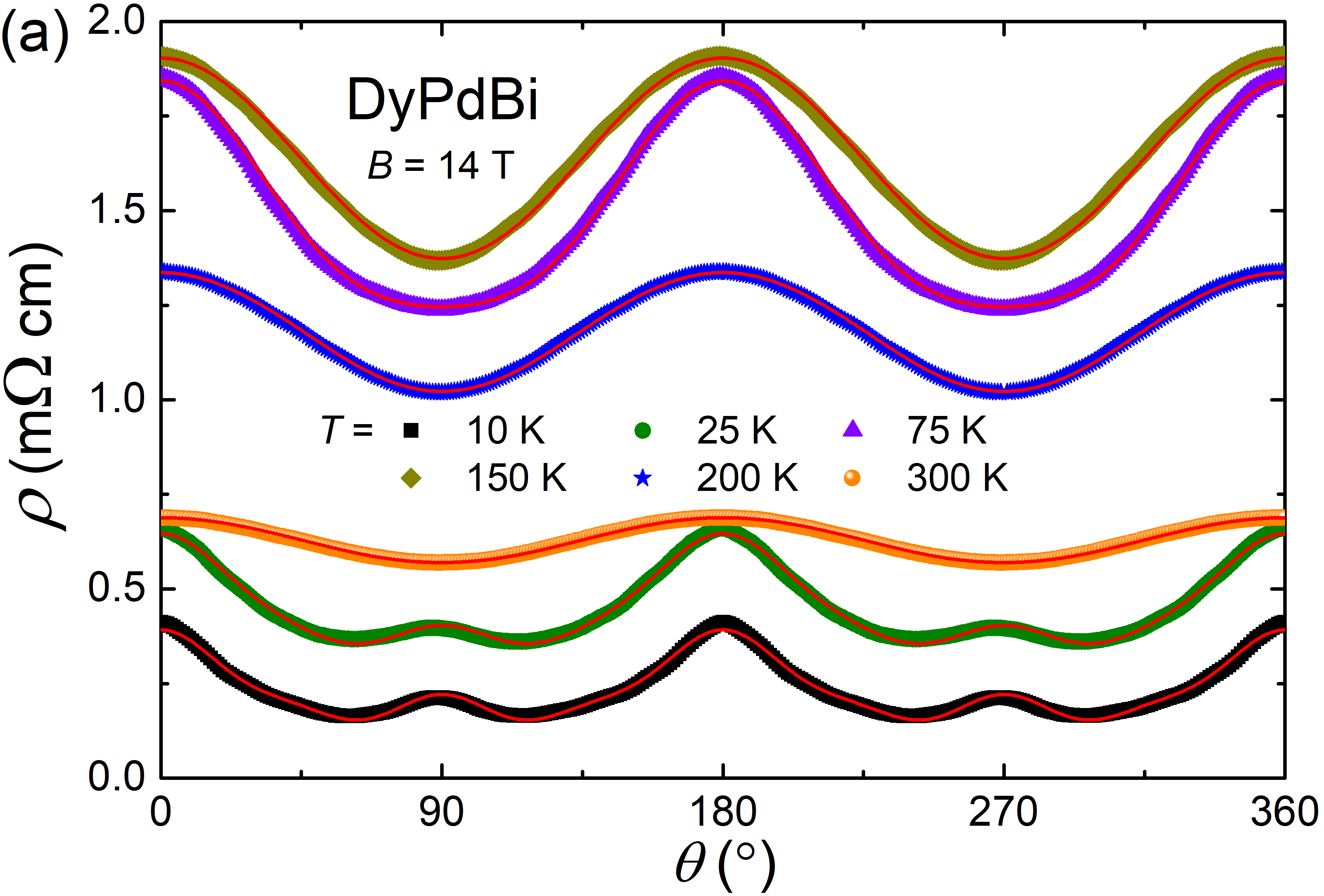}
	\includegraphics[width=0.49\textwidth]{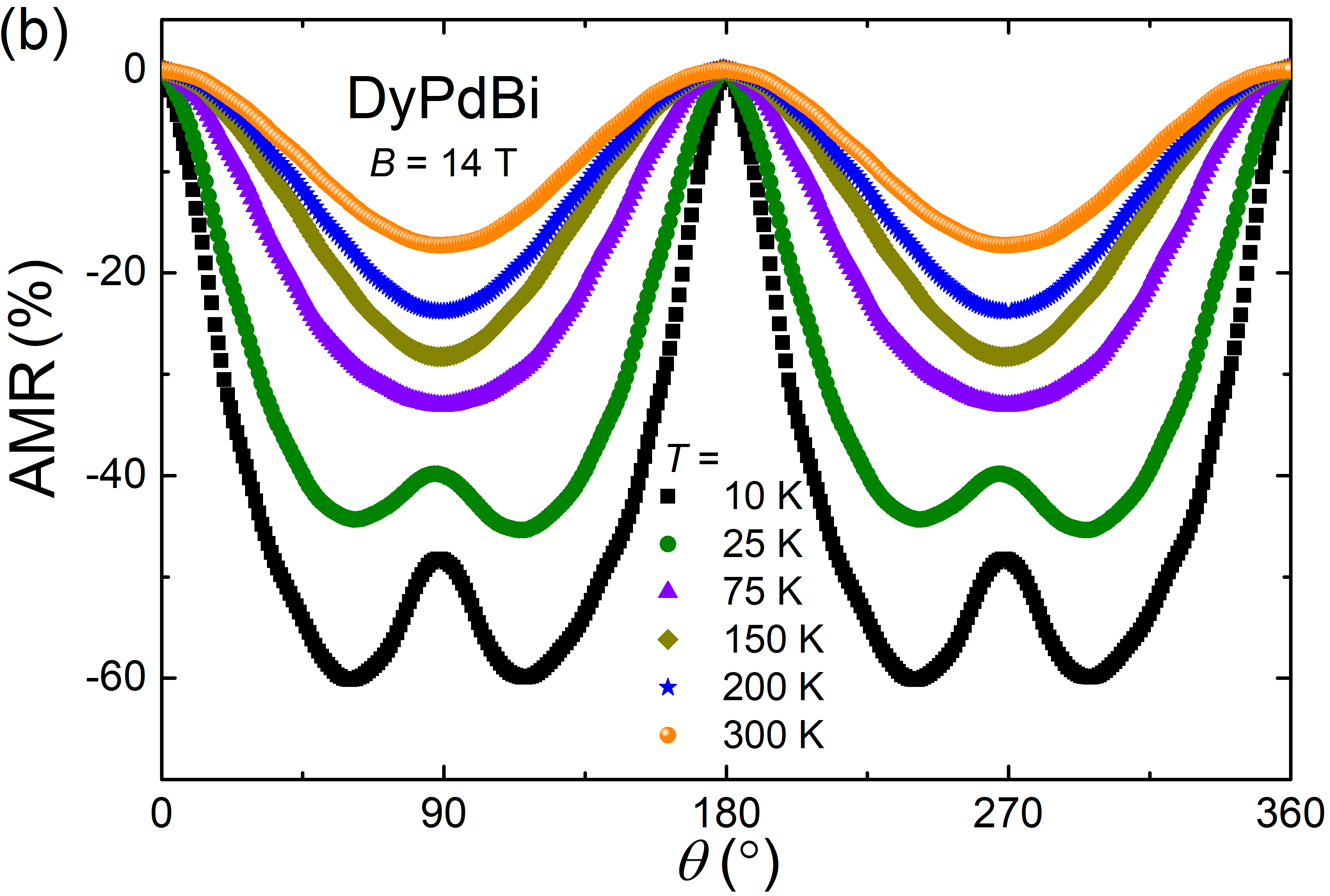}
	\caption{(a) Angular dependence of the electrical resistivity of single-crystalline DyPdBi measured with $j \parallel a$-axis at several different temperatures in $B=14$\,T. Red lines correspond to the fits with Eq.~\ref{Eq_AMR_series}. (b) AMR isotherms derived from data shown in panel (a).
		\label{AMR_plot}}
\end{figure*}
\subsection{Angle dependent magnetoresistance}
The difference between LMR and TMR of DyPdBi is well reflected by AMR=$(\rho_{xx}(\theta)-\rho_{\bot})/\rho_{\bot}$ plotted against field angle $\theta$ in Fig.~\ref{AMR_plot}(b). AMR was calculated from resistivity data collected in magnetic field of 14\,T, rotating from transverse towards parallel direction with respect to the current, at several temperatures, shown in Fig.~\ref{AMR_plot}(a), with $\rho_{\bot}=\rho_{xx}(\theta=0)$. 

Variations we observe are the strongest for the lowest temperature (AMR=-60\% at 10\,K). They weaken with increasing temperature but even at $T=300$\,K AMR attains about -17\%.
  
In polycrystalline materials exhibiting AMR, resistivity varies with field angle in the following way:
\begin{equation}\rho_{xx}(\theta)=\rho_{\bot}-\Delta\rho\cos^2\!\theta\, ,\label{EqAMR}\end{equation} 
where $\rho_{\parallel}=\rho_{xx}(\theta=90^\circ)$, and $\Delta\rho=(\rho_{\bot}-\rho_{\parallel})$ \cite{Jan1957}. 

Such $\rho_{xx}(\theta)$ dependence has also been observed for single crystals of materials with extremely strong classical MR$\propto\!B^2$, for example YSb and LuSb, where AMR reached -80\% and -87\%, respectively \cite{Pavlosiuk2016d,Pavlosiuk2017a}.
However, in single-crystalline DyPdBi we observe additional features, most prominent at low temperatures. 
In order to analyze the $\rho_{xx}(\theta)$ dependence in DyPdBi, the following equation was used, taking into account crystalline anisotropy of a crystal with cubic symmetry \cite{Doring1938,Wu2008,DeRanieri2008}:
\begin{equation}
\rho_{xx}(\theta)=\rho_0(1+A_2\cos2\theta+A_4\cos4\theta+A_8\cos8\theta+...).
\label{Eq_AMR_series}\end{equation}
Fits of Eq.~\ref{Eq_AMR_series}, shown by red solid lines in Fig.~\ref{AMR_plot}(a), correspond well to $\rho_{xx}(\theta)$ experimental data. 
Fitting parameters are listed in Table~I. The $\cos2\theta$ term alone perfectly accounts for angular dependence at high temperatures, from 300\,K down to 150\,K (note that it is equivalent to the function containing $\cos^2\!\theta$, mentioned above). At 75\,K the $\cos4\theta$ contribution is necessary to reproduce four-fold-symmetry features in $\rho(\theta)$, while at 25 and 10\,K there are eight-fold  features, well accounted for by the $\cos8\theta$ term in Eq.~\ref{Eq_AMR_series}.
\begin{table}[h]
	\caption{Parameters obtained from the fitting of Eq.~\ref{Eq_AMR_series} to the experimental $\rho(\theta)$ data of DyPdBi. All $A_4$ and $A_8$ values shown as zeros were smaller than $2\times10^{-4}$.}
\begin{ruledtabular}
\begin{tabular}{rcccc}
			$T$ (K) & $\rho_0$ (m$\Omega$\,cm) & $A_2$ & $A_4$& $A_8$\\\colrule
			 10 & 0.234 & 0.364 & 0.245 & 0.065\\
			 25 & 0.453 & 0.267 & 0.139 & 0.019\\
			 75 & 1.496 & 0.199 & 0.032 & 0\\
			150 & 1.638 & 0.162 & 0 & 0\\
			200 & 1.182 & 0.133 & 0 & 0\\
			300 & 0.629 & 0.094 & 0 & 0\\
		\end{tabular}
	\end{ruledtabular}
\label{AMR_param}
\end{table}

Similarly complex behavior of $\rho(\theta)$ has been found for thin films of ferromagnetic semiconductors \cite{Wu2008,DeRanieri2008}, and most recently also for single-crystalline GdPtBi \cite{Kumar2018a}.
\begin{figure}
	\centering
	\includegraphics[width=0.99\textwidth]{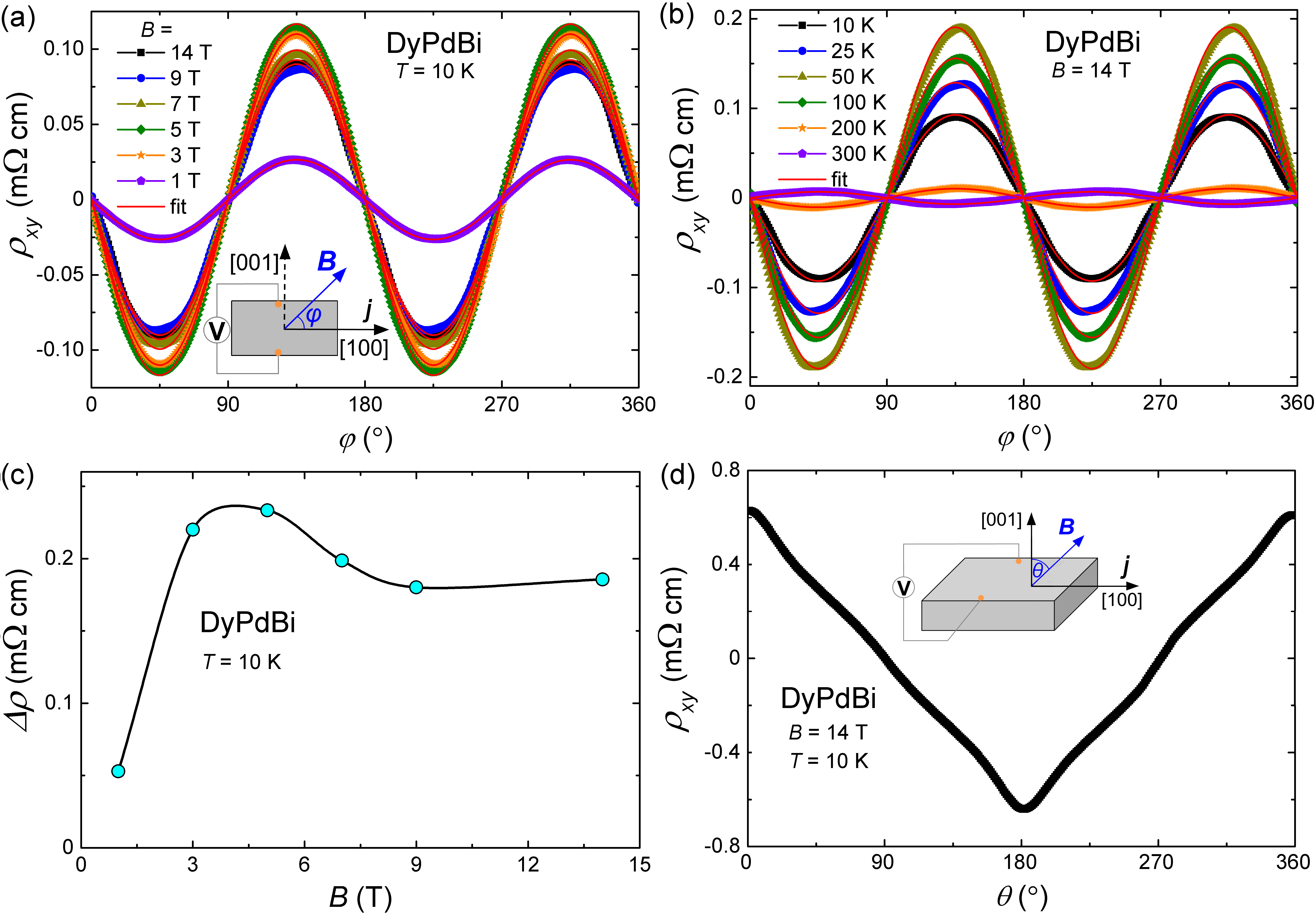}
	\caption{(a) Angular dependence of the planar Hall effect measured for single-crystalline DyPdBi at $T$ = 10\,K in several constant magnetic fields. The measurement geometry is shown in the inset. (b) Angular dependence of PHE in single-crystalline DyPdBi measured as in panel (a) in constant magnetic field $B$ = 14\,T at several different temperatures. Red solid lines in both (a) and (b) represent the fits of the function $\rho_{xy}=\Delta\rho\sin(\varphi)\cos(\varphi)$. (c) Field dependence of $\Delta\rho$ resistivity anisotropy of DyBdBi obtained from the fits shown in panel (a). (d) Hall resistivity of single-crystalline DyPdBi measured as a function of magnetic field direction at $T=10$\,K and in $B=14$\,T. The measurement geometry is shown in the inset.}
	\label{PHE}
\end{figure}
\subsection{Planar Hall effect}
Figure~\ref{PHE} presents the results of PHE measurements performed for DyPdBi with magnetic field rotated in the $x\!-\!y$ plane spanned by perpendicular directions of the current, $j_x$, and of the voltage drop, $V_y$ (see the inset to Fig.~\ref{PHE}(a)).

\begin{figure}[h]
	\includegraphics[width=0.48\textwidth]{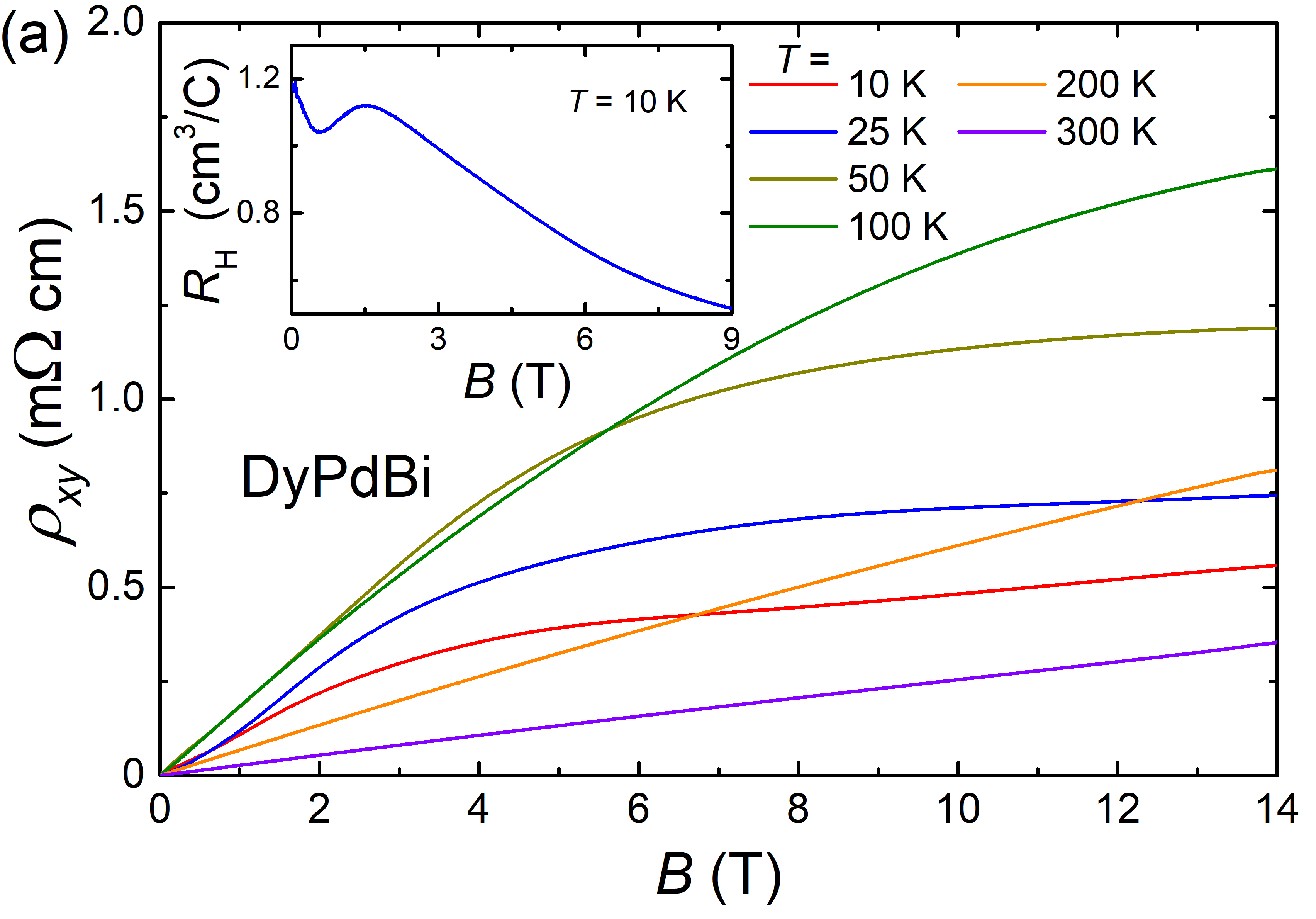}
	\includegraphics[width=0.48\textwidth]{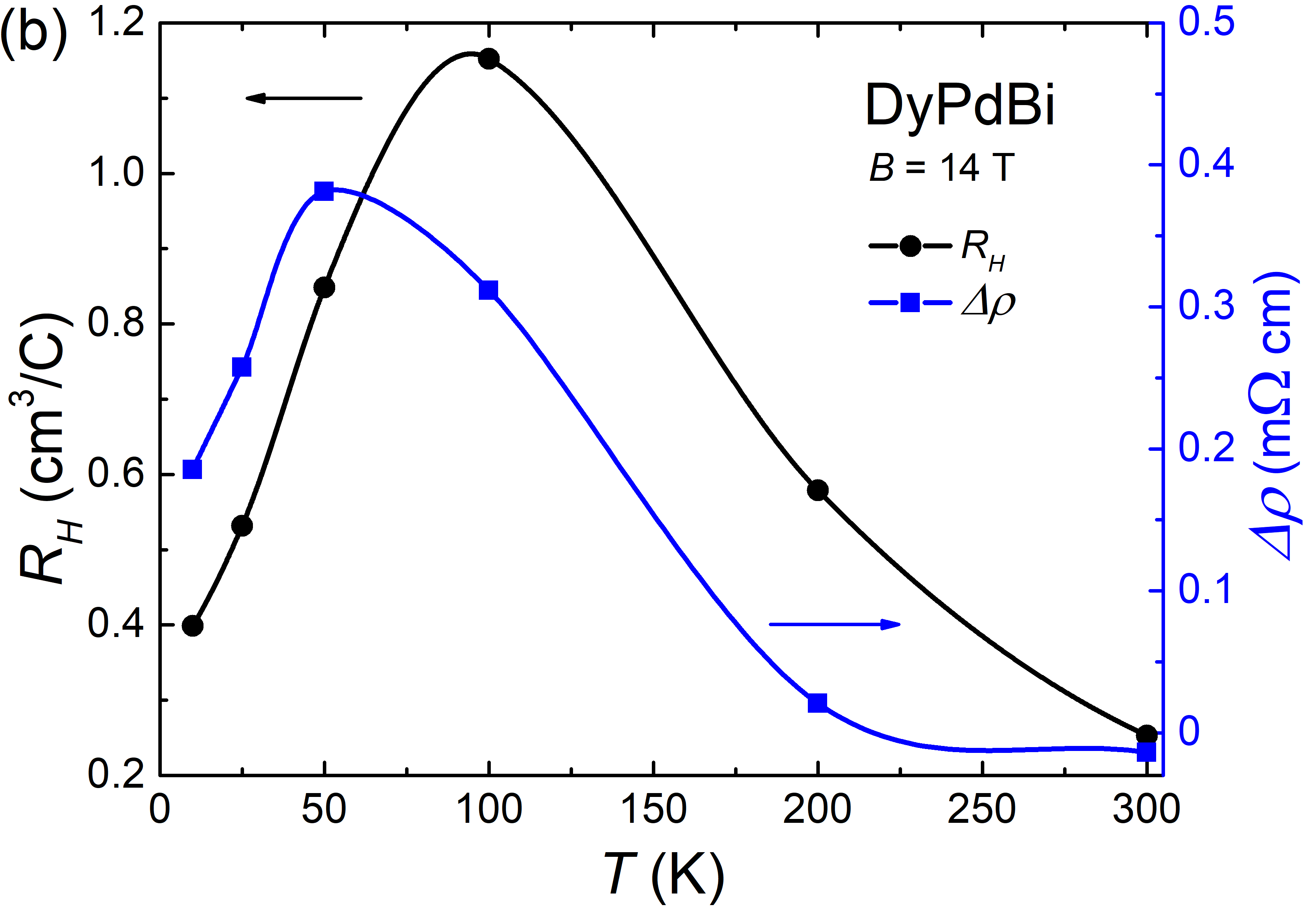}
	\caption{(a) Hall resistivity of single-crystalline DyPdBi measured as a function of magnetic field at several different temperatures with the experimental geometry displayed in Fig.~\ref{PHE}(c). Inset shows the temperature dependence of the Hall coefficient at $T=10$\,K. 
	(b) Comparison of temperature dependence of the Hall coefficient $R_H$, and that of $\Delta\rho(T)$ parameter of PHE, both determined in magnetic field of 14\,T.
	}\label{HE}
\end{figure}

Dependence of PHE resistivity on the angle between current and magnetic field, $\varphi$, should be the same for topological semimetals as for conventional materials  \cite{Jan1957,Burkov2017}: 
\begin{equation}\rho_{yx}(\varphi)=\Delta\rho \sin(\varphi)\cos(\varphi).
\label{PHEq}\end{equation}
Indeed, $\rho_{yx}(\varphi)$ in DyPdBi follows this dependence in the whole range of magnetic field strength and temperature covered by our measurements, as shown in Figs.~\ref{PHE}(a-b). 
Results of fitting Eq.~\ref{PHEq} to the experimental data are shown as red solid lines in Figs.~\ref{PHE}(a) and \ref{PHE}(b). Values of $\Delta\rho$ obtained as parameters of these fittings are plotted as functions of magnetic field and temperature, in Fig.~\ref{PHE}(c) and Fig.~\ref{HE}(b), respectively. Both these dependences are nonmonotonic, in contrast to the behavior of $\Delta\rho$ reported for GdPtBi, where $\Delta\rho$ increased with magnetic field and decreased with temperature \cite{Kumar2018a}, or that observed for Bi$_{2-x}$Sb$_x$Te$_3$, with $\Delta\rho$ also increasing with magnetic field \cite{Taskin2017c}. However, in Bi$_{2-x}$Sb$_x$Te$_3$ $\Delta\rho$ was shown to strongly depend on the Fermi level position and, in consequence, on the density of states at that level, as predicted for CMA \cite{Burkov2017}. Below we show how the nonmonotonic behavior of $\Delta\rho(T)$ in DyPdBi may also have such origin. 

Voltage-current configuration in PHE measurement is the same as in conventional Hall effect, however the latter is measured in field perpendicular to both current and voltage drop. When the field is rotated from that direction towards current, $\rho_{xy}(\theta)$ varies approximately as $\cos(\theta)$, in a manner completely different from that in PHE. This is visualized in Fig.~\ref{PHE}(d) for $\rho_{xy}$ of DyPdBi measured at $T=10$\,K in magnetic field $B=14$\,T. 
%
\subsection{Conventional Hall effect}
For GdPtBi theoretical calculations have shown, that its electronic structure is strongly affected by applied magnetic field \cite{Hirschberger2016a,Suzuki2016}.
Also for DyPdBi calculated spin-polarized band structure is fairly complex, with several bands crossing the Fermi level \cite{Krishnaveni2015}, which makes its carrier concentration sensitive to temperature and magnetic field. 

We thus examined (conventional) Hall effect in DyPdBi and found irregular dependence of its Hall resistivity, $\rho_{xy}$, on temperature and field, as displayed in Fig.~\ref{HE}(a). This implies strong dependence of the electronic band structure of DyPdBi on both $B$ and $T$. 
 
In single-band approximation the Hall coefficient, $R_{\rm H} =\rho_{xy}/B$, directly reflects the carrier concentration $n_{\rm H}=1/(eR_{\rm H})$ ($e$ is elementary charge).
As we show in inset to Fig.~\ref{HE}(a), $R_{\rm H}$ in DyPdBi is strongly and nonmonotonically dependent on the magnetic field strength. The observed  behavior of $R_{\rm H}(B)$ can stem from nonlinear dependence of the carrier concentration on $B$. This contrasts with  GdPtBi, where $\rho_{xy}(B)$ is linearly proportional to $B$ (Ref.~\onlinecite{Shekhar2016a}). 

Temperature dependence of the Hall coefficient of single-crystalline DyPdBi recorded in $B=14$\,T is shown in Figure~\ref{HE}{b}. $R_{\rm H}$ strongly increases with increasing $T$ up to about $T=100$\,K, and then it starts to decrease, returning at 300\,K close to a value observed at 10\,K. The overall $R_{\rm H}(T)$ variation is very similar to the behavior of $\Delta\rho(T)$, as shown in the same Fig.~\ref{HE}(b).

Nonmonotonic $\Delta\rho(T)$ dependence in DyPdBi apparently correlates to temperature variation of carrier concentration, which is in accord with the theoretical prediction  of $\Delta\rho$ being proportional to the density of states at the Fermi level \cite{Burkov2017}. However, the carrier concentration in GdPtBi has been shown to increase monotonically with temperature \cite{Hirschberger2016a}, and monotonic behavior of its $\Delta\rho(T)$ has been reported in Ref.~\onlinecite{Kumar2018a}. 

Thus in both DyPdBi and GdPtBi variations of $\Delta\rho$ with changing $B$ and $T$ are well correlated with behavior of their carrier concentrations. 
\section{Discussion} 
DyPdBi is in many aspects very similar to established Weyl semimetal GdPtBi \cite {Hirschberger2016a}. 
It displays remarkable strong nLMR, AMR and PHE, all hallmarked as fingerprints of CMA, which in turn is a phenomenon specific to Weyl semimetals.
Negative LMR, however, may have sources other than chiral anomaly, and they need to be  discussed before DyPdBi can be pronounced as Weyl semimetal.

Distorted current paths due to conductivity fluctuations induced by macroscopic disorder may cause strong nLMR, as observed in bulk samples of polycrystalline Ag$_{2\pm\delta}$Se (LMR$\approx$-50\% at 110\,K and 14\,T) \cite{Hu2007}.
In DyPdBi nLMR is even stronger. However, strong negative LMR in Ag$_{2\pm\delta}$Se is associated with the large and positive TMR. Moreover, LMR becomes positive when temperature decreases below 50\,K. Additionally, huge spatial inhomogeneity of conductivity in a sample was required to model such a strong nLMR \cite{Hu2007}. These features are exactly opposite to those of DyPdBi where LMR becomes more negative with decreasing temperature, is accompanied by large but negative TMR, and samples were  good-quality single crystals. Therefore macroscopic-disorder scenario of nLMR is not suitable for DyPdBi. 

DyPdBi shows nLMR in paramagnetic state, in weak magnetic fields and high temperatures. This allows us to dismiss the mechanisms of nLMR requiring ferromagnetic order (e.g. see Ref.~\onlinecite{VonHelmolt1993}) or quantum limit (i.e. when $\omega\tau\gg1$, where $\omega$ is the cyclotron frequency and $\tau$ is the transport relaxation time) \cite{Zhang2016a}. 

It is well known that strong apparent nLMR induced by so called current jetting,  can be observed in any material with high-mobility charge carriers \cite{Pippard2009, Liang2018b, Reis2016}. This spurious effect stems from non-uniform current injection in the sample through small contacts. In order to estimate the influence of current jetting on results of our measurements we carried out so called "squeeze test" \cite{Liang2018b}, comparing LMR measured for different positions of voltage contacts. Results of this test clearly show that current jetting has marginal effect on nLMR in our DyPdBi samples (cf Supplemental Material \cite{SupMat}). 

Negative TMR has also been found in HH antimonides \cite{Karla1998} and in HoPdBi \cite{Pavlosiuk2016a}, and attributed to the reduction of spin-disorder scattering  due to alignment of magnetic moments in field, described by de Gennes and Friedel \cite{DeGennes1958}. Resistivity due to spin-disorder scattering in DyPdBi, $\rho_{sd}(B\!=\!0)\!\approx\!0.35\,$m$\Omega$cm, can be roughly estimated from a drop of resistivity occurring when our samples order antiferromagnetically below 3.7\,K. Generally, $\rho_{sd}$ of paramagnets is considered independent from temperature, therefore it can be used to approximate contribution of de Gennes--Friedel mechanism to MR at all temperatures above $T_{\rm N}$. Zero field resistivity of DyPdBi at 10\,K, $\rho_{xx}=1.48\,$m$\Omega$cm, so even complete suppression of the spin-disorder scattering would result in MR$\approx\!-24\%$ (much weaker than $\approx\!-80\%$ we observed). In order to better estimate that effect we measured magnetization, $M$, of DyPdBi at 10\,K in fields up to 7\,T, fitted it with Brillouin function and after extrapolating to the field of 14\,T, found $M=9\mu_{\rm B}$/Dy (as shown in Supplemental Material \cite{SupMat}), less than theoretical saturation value for Dy$^{+3}$ ion, $M_s=10\mu_{\rm B}$.  
Since $\rho_{sd}(B)$ should decrease in applied field as $(1-(M(B)/M_s)^2)$ ~~\cite{DeGennes1958,Karla1998}, we estimate MR due to its reduction as $\approx\!-19\%$.   
Therefore, de Gennes-Friedel mechanism can account for only a fraction of negative MR in DyPdBi. This applies to both LMR and TMR, because de Gennes--Friedel mechanism is isotropic in paramagnets. 
 
Weak localization (WL) is a positive correction to the resistivity, resulting from constructive quantum interference of coherent electron waves upon scattering events forming closed loops. The electron waves have to be in phase to enhance back-scattering probability. The application of a magnetic field dephases the carriers, suppressing WL, which leads to negative MR. WL is prominent in low-mobility materials with strong spin-orbit coupling, such as DyPdBi. Magnetoconductivity resulting from WL  
is anisotropic in regard to applied magnetic field. 

%
\begin{figure}[h]
	\includegraphics[width=0.6\textwidth]{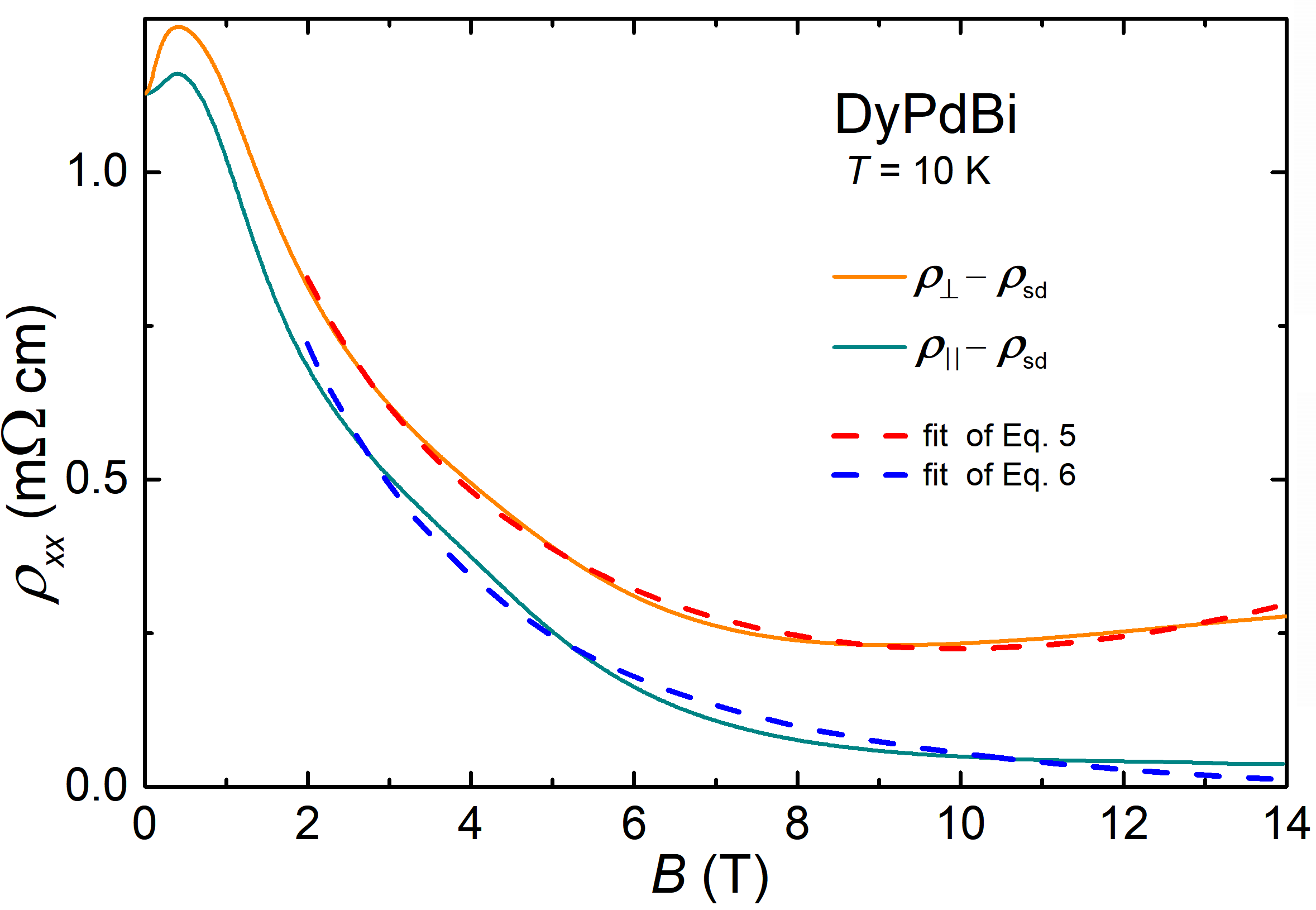}
	\caption{Resistivity measured at $T=$10\,K, in transverse and longitudinal fields, with the spin-disorder (de Gennes-Friedel) resistivity subtracted. Dashed lines represent the fits with Equations \ref{rhoperp} and \ref{rhopar}.}
	\label{CMA_WL_rho} 
\end{figure}

Formula to fit magnetoconductivity arising from the weak localization in three dimensions has been proposed in Ref.~\onlinecite{Lu2017b}:
\begin{equation}\delta\sigma^{qi}_{xx}(B)=c_1\frac{B^2\sqrt{B}}{B^2+B_c^2}+c_2\frac{B_c^2B^2}{B^2+B_c^2}.\end{equation}
We incorporated that formula in equations describing resistivities, shown in Fig.~\ref{CMA_WL_rho}, obtained after subtraction of isotropic $\rho_{sd}(B)$ from $\rho_{\bot}(B)$ and $\rho_{\parallel}(B)$:  
\begin{equation}
\rho_{\bot}(B)-\rho_{sd}(B)=\rho_0+\delta\sigma^{qi}_{xx}(B)^{-1}+c_{\rm D}B^2,~~
\label{rhoperp}\end{equation}
\begin{equation}
\rho_{\parallel}(B)-\rho_{sd}(B)=\rho_0+(\delta\sigma^{qi}_{xx}(B)+c_{\rm W}B^2)^{-1}.\label{rhopar}
\end{equation}
These equations contain additional terms: Drude resistivity $c_{\rm D}B^2$, for $\rho_{\bot}(B)$, and the $c_{\rm W}B^2$ magnetoconductivity term, expected for CMA, for $\rho_{\parallel}(B)$\cite{Son2013,Burkov2017,Zhang2016a}. 
We obtained satisfactory fits of equations~\ref{rhoperp} and \ref{rhopar} to experimental data, as shown in Fig.~\ref{CMA_WL_rho}. The Drude term accounts for a slight upturn of $(\rho_{\bot}-\rho_{sd})$ in fields above 9\,T, whereas the $c_{\rm W}B^2$ contribution magnetoconductivity is essential in fitting to $(\rho_{\parallel}-\rho_{sd})$. Therefore, even though both, de Gennes--Friedel effect and weak localization lead to negative MR in DyPdBi, we may assume that chiral anomaly also contributes to nLMR in this compound. 
\section{Conclusions}
The results obtained for high-quality single crystals of DyPdBi revealed strong negative longitudinal magnetoresistance, but also somehow weaker but also negative transverse magnetoresistance. They result in giant anisotropic magnetoresistance and large planar Hall effect. All these phenomena can be partially ascribed to the chiral magnetic anomaly, a hallmark feature of topological semimetals. The compound exhibits complex behavior of the magnetoresistance isotherms, which do not saturate up to the highest magnetic fields and attain huge negative values at low temperatures, even in transverse measurement geometry. 
Even at room temperature, AMR of DyPdBi has very large negative values in strong magnetic fields. The amplitude of PHE is also large and depends on temperature and magnetic field in irregular manners, probably due to strongly $T$- and $B$-dependent carrier concentration, as revealed by the Hall effect results. 
Keeping in mind that the results of single-band Drude approach yield only the upper limit of real carrier concentration in a multi-band material like DyPdBi, these dependences of carrier concentration provide explanation of the differences in behavior of the PHE amplitudes for DyPdBi and GdPtBi. 
The obtained data indicate that DyPdBi is a convenient platform for tuning the magnitude of PHE by chemical doping. The presence of AMR and PHE in this material makes it a good candidate for magnetic sensor applications.
\begin{acknowledgments}
Work was financially supported by the National Science Centre of Poland, grant no. 2015/18/A/ST3/00057. 
\end{acknowledgments}
%
%
\renewcommand*{\familydefault}{\sfdefault}
\setcounter{figure}{0}
\renewcommand{\thefigure}{S\arabic{figure}}
\newpage
\noindent\begin{large}Supplemental Material\end{large}\\\\
\begin{large}{\bf Negative longitudinal magnetoresistance as a sign of a possible\\chiral magnetic anomaly in the half-Heusler antiferromagnet DyPdBi}\\\\ 	
Orest Pavlosiuk, Dariusz Kaczorowski and Piotr Wi\'sniewski
\end{large}\\
{\it Institute of Low Temperature and Structure Research,
	Polish Academy of Sciences,\\ P. O. Box 1410, 50-950 Wroc{\l}aw, Poland}\\
\vspace{-1cm}\subsection*{Laue diffraction.}
\begin{figure}[h]
	\includegraphics[width=0.6\textwidth]{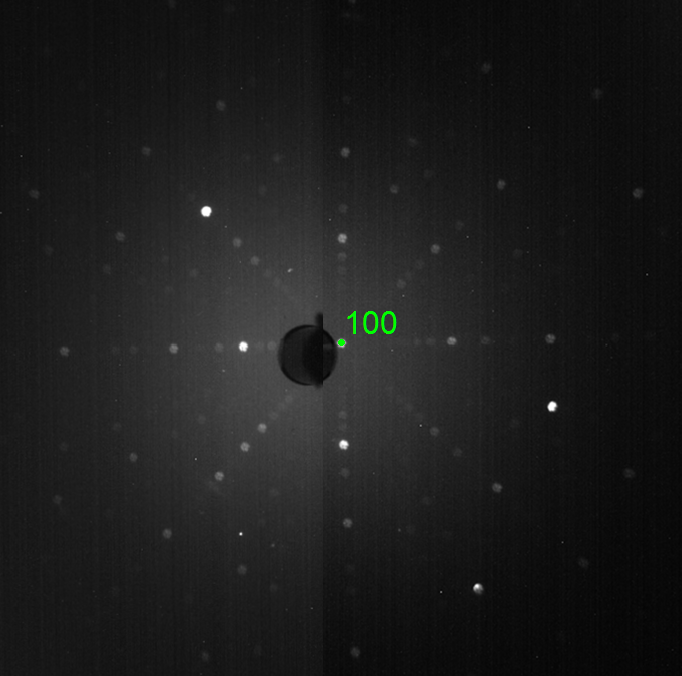}
	\caption{Laue diffraction pattern of DyPdBi single crystal taken with incident X-ray beam close to the [100] crystallographic direction.}
		\label{Laue_img}
\end{figure}

High quality of the single crystal studied was confirmed by backscattering Laue method using a Proto-COS Laue system. An example of the X-ray Laue diffraction pattern is shown in Fig.~\ref{Laue_img}.
\subsection*{Longitudinal magnetoresistance measured with different position of voltage contacts -- "squeeze test" of current jetting.}
\begin{figure}[b]
	\includegraphics[width=\textwidth]{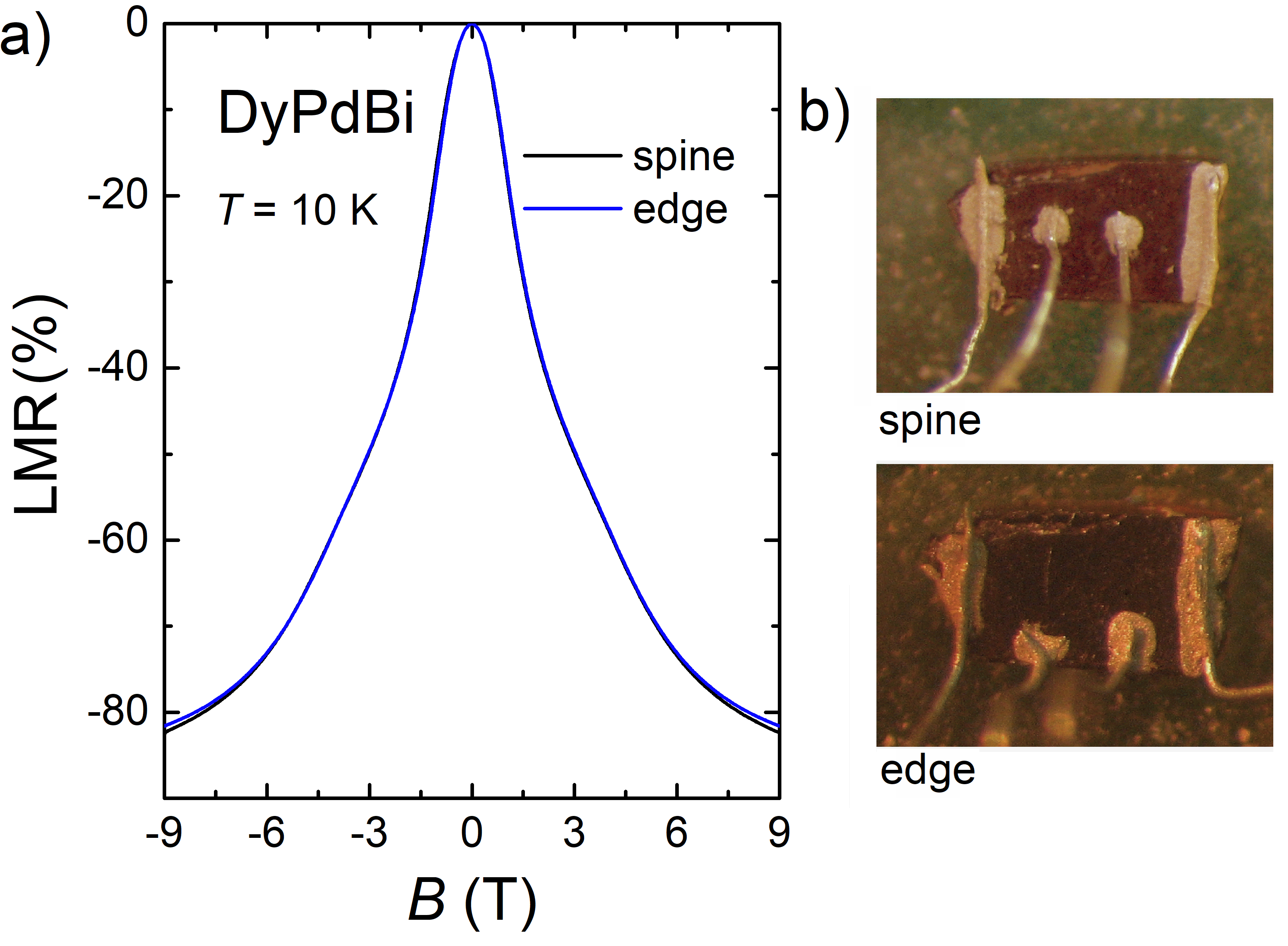}
	\caption{Longitudinal magnetoresistance measured for two different position of voltage contacts.}
		\label{Squeeze_img}
\end{figure}

\vspace{-0.5cm}
In order to estimate the influence of current jetting on results of our measurements we carried out so called "squeeze test" [1]. LMR was measured on the same sample, with different positions of voltage contacts, denoted as "spine" and "edge", and shown in Fig.~\ref{Squeeze_img}b. For all measurements of magnetoresistance we prepared current contacts stretching for whole width of samples, so as to minimize the current jetting. 

Result of this test is shown in Fig.~\ref{Squeeze_img}a. The difference between "spine" and "edge" configurations is about 1\%, which shows that current jetting has marginal effect on LMR in DyPdBi. 
\subsection*{Magnetization of DyPdBi and its fitting with Brillouin function.}
\begin{figure}[h]
	\includegraphics[width=0.6\textwidth]{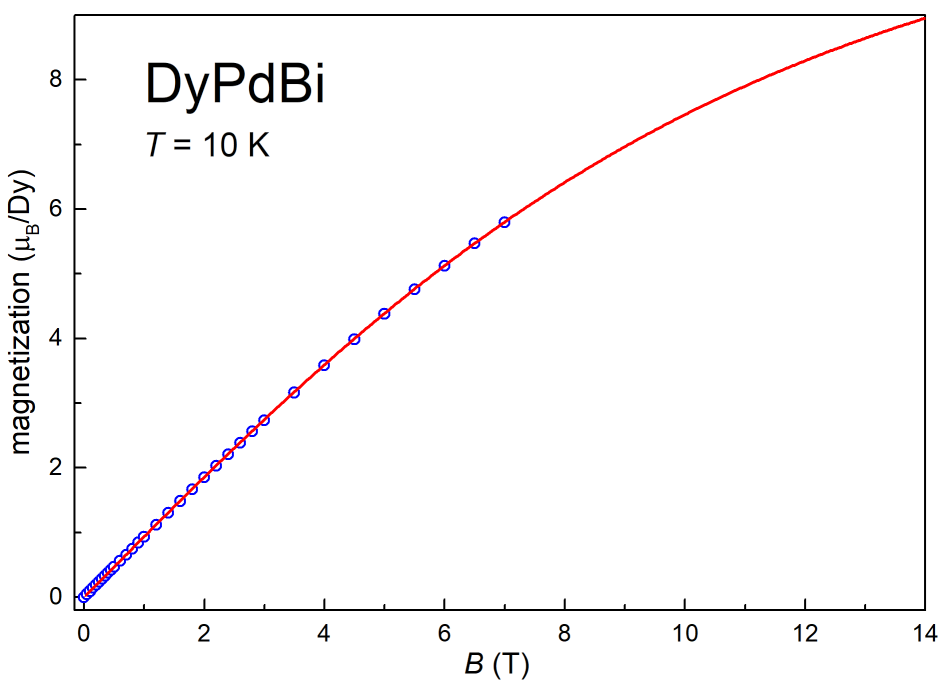}
	\caption{Magnetization of DyPdBi at temperature of 10\,K. Red solid line represents  Brillouin function, which was fitted to data collected in field up to 7\,T.}
		\label{Magnetiz}
\end{figure}

\vspace{-0.5cm}
In order to estimate spin-disorder magnetoresistance, described by de Gennes-Friedel mechanism, we measured magnetization of DyPdBi at 10\,K, as shown in Fig.~\ref{Magnetiz}. Next we performed fit of Brillouin function to collected data. Plotting fitted function in range of field extended up to 14\,T yields for that field magnetization of 9$\mu_{\rm B}$ per atom of Dysprosium, less than 10$\mu_{\rm B}$ - theoretical saturation value for Dy$^{+3}$ ion. This plot also shows that magnetization at 14\,T is still quite far from saturation. 

We used $M(B)$ described by the fitted Brillouin function to calculate spin-disorder resistivity: $\rho_{sd}(B)=\rho_{sd}(B=0)(1-(M(B)/M_s)^2)$ [2, 3], which was then subtracted from resistivities measured in longitudinal and transverse magnetic fields. Results of that subtraction is shown in Fig.~5 of the main text. 
\\
\subsection*{Supplemental references:}\vspace{-0.6cm}
\noindent 1. S. Liang, J. Lin, S. Kushwaha, J. Xing, N. Ni, R. J. Cava, and N. P. Ong,\\ \hspace{2cm} \href {\doibase 10.1103/PhysRevX.8.031002} {Phys. Rev. X {\bf 8}, 031002 (2018)}.\\
2. P. de Gennes and J. Friedel, \href {\doibase 10.1016/0022-3697(58)90196-3}{J. Phys. Chem. Solids {\bf 4}, 71 (1958)}.\\
3. I. Karla, J. Pierre, and R. Skolozdra, \href
  {\doibase 10.1016/S0925-8388(97)00419-2}{J. Alloys and Compd. {\bf 265}, 42 (1998)}.

\end{document}